\documentclass[12pt,aps,prb,preprint]{revtex4}   % prb , style for Physical Review B and AJP are similar\usepackage{amsmath}    % need for 
\usepackage{amsmath}    % need for subequations
\usepackage{amsfonts}  %note how statements can be commented out
\usepackage{graphicx}   % for figures

\usepackage{graphicx}% Include figure files
\usepackage{dcolumn}% Align table columns on decimal point
\usepackage{bm}% bold math

\newcommand{\be}{\begin{equation}}
\newcommand{\ee}{\end{equation}}
\newcommand{\bea}{\begin{eqnarray}}
\newcommand{\eea}{\end{eqnarray}}

\newcommand{\rme}{{\rm{e}}}

\newcommand{\iGA}{{i}}
\newcommand{\jGA}{{j}}

\begin{document}

%\preprint{APS/123-QED}

\title{A simplified approach to electromagnetism using geometric algebra}  
\author{James M.~Chappell}
\email{james.m.chappell@adelaide.edu.au}
\affiliation{(a) School of Electrical and Electronic Engineering, \\
(b) School of Chemistry and Physics, University of Adelaide, South Australia
5005, Australia}
\author{Azhar Iqbal}
\affiliation{a) School of Electrical and Electronic Engineering, University of Adelaide,
South Australia 5005, Australia\\
b) Centre for Advanced Mathematics and Physics, National University of
Sciences \& Technology, Sector H-12, Islamabad, Pakistan}
\author{Derek Abbott}
\affiliation{School of Electrical and Electronic Engineering, University of Adelaide, South Australia 5005, Australia }
\date{\today}

\begin{abstract}
A new simplified approach for teaching electromagnetism is presented using the formalism of geometric algebra (GA) which does not require vector calculus or tensor index notation, thus producing a much more accessible presentation for students.  The four-dimensional spacetime proposed is completely symmetrical between the space and time dimensions, thus fulfilling Minkowski's original vision.
In order to improve student reception we also focus on forces and the conservation of energy and momentum, which take a very simple form in GA, so that students can easily build on established intuitions in using these laws developed from studying Newtonian mechanics.  
The potential formulation is also integrated into the presentation that allows an alternate solution path, as well as an introduction to the Lagrangian approach. Several problems are solved throughout the text to make the implementation clear.  
We extend previous treatment of this area, through including the potential formulation, the conservation of energy and momentum, the generalization for magnetic monopoles, as well as simplifying previously reported results through eliminating the need for the spacetime metric.
\end{abstract}

\pacs{01.55.+b, 41.20.-q} % PACS, the Physics and Astronomy Classification Scheme.
\maketitle

\section{Introduction}

Maxwell's equations were first published in 1865 \cite{Maxwell:1865} mathematically describing then recently discovered electromagnetic phenomena.  His equations were written for three-space, requiring 12 equations in 12 unknowns. These equations were later rewritten by Heaviside and Gibbs, in the newly developed formalism of dot and cross products, which reduced them to the four equations now seen in most modern textbooks  \cite{Griffiths:1999} and shown below in S.I. units
\bea \label{MaxwellClassical}
\mathbf{\nabla} \cdot \mathbf{E} & = & \frac{\rho}{\epsilon},   \,\,\,\,\,\,\, \rm{(Gauss\text{'} \,\, law)} ; \\ \nonumber
\mathbf{\nabla} \times \mathbf{B} - \frac{1}{c^2} \partial_t \textbf{E} & = & \mu_0 \mathbf{J},  \,\,\,  \rm{(Amp\grave{e}re\text{'}s \,\, law)} ; \\ \nonumber
\mathbf{\nabla} \times \mathbf{E} +  \partial_t \mathbf{B}  & = &  0 , \,\,\,\,\,\,\,\,\, \rm{(Faraday\text{'}s \,\, law)} ; \\ \nonumber
\mathbf{\nabla} \cdot \mathbf{B}  & = &  0 , \,\,\,  \,\,\,  \,\,\,  \rm{(Gauss\text{'} \,\, law \,\, of \,\, magnetism)} ,  \\ \nonumber
\eea
where $ \mathbf{E} , \mathbf{B},\mathbf{J} $ are conventional vector fields, with $ \mathbf{E} $  the electric field strength V/m and $ \mathbf{B} $ the magnetic field strength in units of $\rm{Teslas}$ and $ \mathbf{\nabla} $ is the three-vector gradient given by
\bea
\mathbf{\nabla} & = & e_1 \frac{\partial}{\partial x} + e_2 \frac{\partial}{\partial y} + e_3 \frac{\partial}{\partial z} , \\ \nonumber
\eea
and $ e_1, e_2, e_3 $ are the three real Euclidean space orthonormal vectors, with $ e_i \cdot e_j = \delta_{ij} $, where $ \delta_{ij} $ is the Kronecker-delta symbol.
Maxwell's four equations given in Eq.~(\ref{MaxwellClassical}) along with the Lorentz force law
\be \label{LorentzForceLaw}
\mathbf{K} = q (\mathbf{E}+\mathbf{v} \times \mathbf{B}),
\ee
completely summarize classical electrodynamics \cite{Griffiths:1999}.   It can be seen that we have adopted the convention that a three-space vector $ \textbf{v} = v_1 e_1 + v_2 e_2 + v_3 e_3 $, is identified in bold, which is in agreement with ISO 80000-2 international standard for mathematical notation. 

Hestenes in 1966 \cite{Hestenes:1966}, also refer \cite{Hestenes:2003}$^,$\cite{Vold:1993}$^,$\cite{Baylis:1994} in this journal, using the formalism of GA, further condense Maxwell's  four coupled first order differential equations (DE's) into just a single first order DE
\be \label{eq:maxwellGA3DVold}
\left (\frac{1}{c} \partial_t + \mathbf{\nabla} \right )  F = c \mu_0 J,
\ee
where the field strength $ F =  \textbf{E} + \iGA c \textbf{B} $, with the sources held in $ J = c \rho -  \textbf{J} $. The symbols $ F, J $, are called multivectors and as we shall later see can be visualized as containing a combination of scalar, vector, pseudovector and higher order terms.  It might appear invalid, from the perspective of vector calculus, that in the source term $ J $, we have subtracted the vector quantity $ \textbf{J} $ from the scalar quantity $ c \rho $. However this is an unnecessary limitation, which severely limits the power of vectors.  We can also see the benefit in relaxing this restriction, in that the field strength $ F $ can now hold both the electric and magnetic fields in a single variable.  This is analogous to the formalism of complex numbers that 
contain both real and imaginary numbers as a single entity, which is a very successful formalism for two dimensions. See Appendix A to confirm the equivalence of Eq.~(\ref{eq:maxwellGA3DVold}) with Maxwell's four equations.   

This comparison is investigated further in Table~\ref{tableFormalisms} where we compare GA, Gibbs vector calculus, and the tensor calculus formalisms.  Tensor formalism requires two equations, four-dimensional spacetime and a metric in order to represent Maxwell's equations, compared with Eq.~(\ref{eq:maxwellGA3DVold}) in three-space using GA, without the use of a metric.  Tensor notation is briefly described in Appendix B in order to allow a direct comparison. Differential forms can also be used, but still require two separate equations to represent Maxwell's equations.

\begin{table}
	\centering

\begin{tabular}{|l|l|l|l|l|}
\hline
 EM theory  & Gibb's vector calculus & Tensors & GA  \\
\hline \hline
Fields & $ \textbf{E} $,$ c \mathbf{B} $ & $ F^{\alpha \beta} $ &  $ \mathbf{E} + \iGA c \mathbf{B}  $  \\
EM equations & $ \mathbf{\nabla} \cdot \mathbf{E}  = \frac{\rho}{\epsilon} $, $\mathbf{\nabla} \times \mathbf{E} +  \partial_t \mathbf{B}  =  0 $ &  $  F^{\alpha \beta }_{, \alpha} = c \mu_0 J^{\beta} $ &  $ \Box F = c \mu_0 J $ \\
 & $\mathbf{\nabla} \times \mathbf{B} - \frac{1}{c^2} \partial_t \mathbf{E} = \mu_0 \mathbf{J} $, $ \mathbf{\nabla} \cdot \mathbf{B}  =  0 $ &  $ \epsilon^{\alpha \beta \gamma \delta} F_{\alpha \beta , \gamma} = 0 $   &   \\
Charge conservation & $ \mathbf{\nabla} \cdot \mathbf{J} + \frac{\partial \rho }{\partial t} = 0 $ & $ J^{\alpha}_{, \alpha } = 0 $ & $ \Box \cdot J = 0 $ \\
Energy in fields $U$ & $ \frac{1}{2} \epsilon ( \mathbf{E}^2 + c^2 \mathbf{B}^2 ) $ & $  \epsilon (F^{0 \gamma} F^{0}_{\gamma} + \frac{1}{4} F_{\mu \nu}  F^{\mu \nu} ) $ &  $ -\frac{1}{2} \epsilon F F^{\dagger} $  \\
Poynting vector $\mathbf{S}/c$ & $ \frac{1}{\mu_0 c} (\mathbf{E} \times \mathbf{B}) $ &  $ \epsilon F^{0 \gamma} F^{j}_{\gamma}  $ &   \\
Invariants & $ c^2 \mathbf{B}^2 - \mathbf{E}^2 $ & $ \frac{1}{2} F^{\alpha \beta} F_{\alpha \beta} $ &  $ F^2 $  \\
         & $ \frac{2}{c} \mathbf{B} \cdot \mathbf{E} $ & $ \frac{1}{4} \epsilon_{\alpha \beta \gamma \delta} F^{\alpha \beta} F^{\gamma \delta} $ &   \\
Action $ E_p - E_k $ & $ \rho V - \textbf{J} \cdot \mathbf{A} $ & $ J^{\alpha} A_{\alpha} $ &  $ J \cdot A $  \\
Minkowski Force & $ \mathbf{K} = \gamma q(\mathbf{E}+\mathbf{v} \times \textbf{B}) $ & $ c K^{\mu} = q v_{\nu} F^{\mu \nu } $ &  $ c K =  - q \underline{v} F  $  \\
Conservation energy & $ \frac{\partial U }{\partial t} + \mathbf{\nabla} \cdot \mathbf{S} = -\textbf{J} \cdot \mathbf{E}  $ & $ c \partial_{\beta} T^{\alpha \beta} = - g^{\alpha \rho} F_{\rho \gamma } J^{\gamma} $ &  $ c \Box \cdot T_0 = J F $  \\
momentum $  \mathbf{S}/c^2 $ & $ \mathbf{\nabla} \cdot \stackrel{\leftrightarrow}{T}  - \frac{1}{c^2} \frac{\partial \mathbf{S} }{\partial t} = \rho \textbf{E} + \mathbf{J} \times \mathbf{B} $ &  &  $ c \Box \cdot T_i = (J F)_i $  \\
Potential function $A$ & $ \mathbf{E} = - \mathbf{\nabla} V - \frac{\partial \mathbf{A}}{\partial t}, \,\, \mathbf{B} = \mathbf{\nabla} \times \textbf{A} $ &  $  F_{\alpha \beta } = A_{\alpha,\beta} -A_{\beta,\alpha} $ &  $ F = \Box A $ \\
Lorenz gauge & $ \lambda = \mathbf{\nabla} \cdot \mathbf{A} + \frac{1}{c^2} \frac{ \partial V}{\partial t} = 0 $ & $ A^{\alpha}_{, \alpha } = 0 $ &  $ \Box \cdot A = 0 $  \\
Maxwell's Equations & $ \Box^2 \mathbf{A} = - \mu_0 \mathbf{J}, \, \Box^2 V = - \frac{\rho}{\epsilon_0} $ & $ \Box^2 A^{\mu} = - c \mu_0 J^{\mu} $ &  $ \Box^2 A = c \mu_0 J $  \\
Stress energy tensor & $ \stackrel{\leftrightarrow}{T} $ (See below for definition) & $ F^{\alpha \gamma} F^{\beta}_{\gamma} + \frac{1}{4} g^{\alpha \beta} F_{\gamma \nu}  F^{\gamma \nu}  $ &  $ T_{\mu} = \frac{1}{2} F n_{\mu} F $  \\
 
\hline
\end{tabular}
\caption{Comparison of mathematical formalisms, showing relative simplicity of mathematical expressions in GA. Note: The blanks in several rows in the column for GA is because the relevant equation is automatically part of the equation above. For example, there is a blank row below the $ F^2 $ entry for GA, because the separate scalar and pseudoscalar components of this expression produce the two invariants in the vector column. The Maxwell stress energy tensor given by $ \stackrel{\leftrightarrow}{T} \leftrightarrow T_{ij} = \epsilon_0 (E_i E_j -\frac{1}{2} \delta_{ij} E^2) + \frac{1}{\mu_0} (B_i B_j -\frac{1}{2} \delta_{ij} B^2) $. We have also the four-vector $ \underline{n} $ defined as, $ n_0 = e_0 $ and $ n_i = e_i e_0 $. }    \label{tableFormalisms}
\end{table}

Even though Gibbs was able to reduce Maxwell's twelve equations down to four, as mentioned, his formalism for vectors had significant structural limitations.  For example, the cross product only applies in three dimensional space, because in four dimensions there is an infinity of perpendicular vectors. However, probably most serious in terms of students learning physics, is that, conventional vectors do not integrate with established algebraic intuitions regarding basic operations.  That is, there is no division operation, the cross product does not apply in two dimensions and one cannot freely add vectors to previously known algebraic elements (scalars), so that vector algebra becomes a monolithic structure unto itself.  Hence the intuitive understanding of physics concepts, as well as general geometric understanding, which depends on the understanding of vectors, is significantly reduced.

Historically, as vectors became more popular in physics and in various other fields, new scientific discoveries such as quantum mechanics and relativity meant that vector analysis needed to be supplemented by a basket of other mathematical techniques such as: tensors, spinors, matrix algebra, Hilbert spaces,
differential forms etc. As noted in \cite{Simons:2009}, `The result is a bewildering plethora of mathematical techniques which require much learning and teaching, which tend to fragment the subject and which embody wasteful overlaps and requirements of translation'.

Alternatively, the mathematical formalism of geometric algebra, developed by Clifford in 1878, as convincingly argued by Hestenes \cite{GA2},  provides a single unified mathematical language for physics.  Clifford's development of GA was based on work by Grassman on the exterior product and Hamilton's work on quaternions from 1843.  With this formalism we can freely multiply and divide by vectors in a natural way as well as adding and subtracting diverse elements such as scalars and vectors.  This is achieved primarily through combining the dot product and cross product into a single new product called the geometric product, which for two vectors $ u $ and $ v $ is defined as
\be \label{geometricProductuv}
u v = u \cdot v + \iGA u \times v = u \cdot v + u \wedge v.
\ee
This simple device (which works in any number of dimensions) allows Maxwell's equations to be seamlessly melded together as shown in Eq.~(\ref{eq:maxwellGA3DVold}). Maxwell himself adopted quaternions in his follow up work in 1873, \itshape{Treatise on Electricity and Magnetism} \upshape \cite{Maxwell:1873}, writing his complete set of equations in this formalism.  However Maxwell died in 1879 just after Clifford developed GA and so never had an opportunity to implement Clifford's approach, which subsumes quaternions into a complete algebraic structure. 
Maxwell's equations however, have stood the test of time and despite the fact that it is nearly 150 years since he first presented his equations, his basic premises and equations have remained valid.
It has been pointed out \cite{Tu2005}, for example, that Maxwell's equations assume a massless photon.  If photons were not massless an extra current term would need to be added to the right hand side of his equations, but many precision experimental tests on the photon's mass have confirmed a mass approaching zero. Also Maxwell's third and fourth equations, Amp$\grave{\rm{e}}$re's law and Gauss's law of magnetism, are a way of stating that magnetic monopoles do not exist and in fact despite diligent searches, their existence has not been verified \cite{Griffiths:1999}. Although this is consistent with special relativity, which asserts that $ \textbf{B} $ fields are created by relative motion, being demonstrated in \cite{Lorraine1988}, by using electrostatics and relativity alone one can derive the Lorentz force law, Eq.~(\ref{LorentzForceLaw}). 

Ironically, also, Maxwell's equations developed forty years before Einstein's 1905 relativity paper, satisfy the relativity principle even though they were developed under the assumption of an aether.
Hence, despite various attempts at modifications, Maxwell's original equations remain essentially intact.

\subsection{A description of the GA formalism}

The mathematical elements of geometric algebra (GA) are constructed from real scalars and unit vectors, with all other algebraic quantities being derived using the outer product.
For example, for a three dimensional space we have the orthonormal basis vectors $ e_1, e_2 $ and $ e_3 $, satisfying $ e_{i} \cdot e_{j}=\delta _{ij} $, with which we produce the bivectors  $ e_1 \wedge e_2 $, $ e_2 \wedge e_3 $, $ e_1 \wedge e_3 $ and the trivector  $  e_1 \wedge e_2 \wedge e_3$. All algebraic elements can be freely added together to form a quantity called a multivector.

Elements of the algebra form a linear space of multivectors over the real numbers.
That is, if  $ M_1 $ and $ M_2 $ are multivectors, then so is  $ a_1 M_1 + a_2 M_2 $	for any real numbers $ a_1 $  and $ a_2 $. 

Multiplication between algebraic elements is defined to be the geometric product, which for two vectors $ u $ and $ v $ is shown in Eq.~(\ref{geometricProductuv}), where $u \cdot v$ is the conventional symmetric dot product and $%
u \wedge v $ is the anti-symmetric outer product related to the Gibb's cross product by $%
u \times v=-\iGA u \wedge v $ or $ u \wedge v = \iGA u \times v $, where $\iGA =e_{1}e_{2}e_{3}$. 

The naturalness of this definition for the geometric product can be seen by using the distributive law to expand the product of two vectors
\bea \label{VectorProductExpand}
& & (e_1 u_1 + e_2 u_2 + e_3 u_3 ) ( e_1 v_1 + e_2 v_2 + e_3 v_3 ) \\ \nonumber
& = & u_1 v_1 + u_2 v_2 + u_3 v_3 + (u_2 v_3 - v_2 u_3 ) e_2 e_3 + (u_1 v_3 - u_3 v_1 ) e_1 e_3 + (u_1 v_2 - v_1 u_2 ) e_1 e_2 \\ \nonumber
& = & \textbf{u} \cdot \textbf{v}  + \textbf{u} \wedge \textbf{v}  = \textbf{u} \cdot \textbf{v}  + i \textbf{u} \times \textbf{v}, \\ \nonumber
\eea
using $ e_i^2 = 1 $ and $ e_i e_j = - e_j e_i $, which thus provides a definition of the dot and outer products in component form. This also shows how the basic distributive law for expanding brackets naturally splits into dot and outer products, thus integrating these with basic algebra. For a detailed calculation comparing the cross and the outer product using GA, please refer to Appendix~C. 

For distinct basis vectors we find from Eq.~(\ref{geometricProductuv})
\begin{equation}
e_{i}e_{j}=e_{i} \cdot e_{j}+e_{i}\wedge e_{j}=e_{i}\wedge e_{j}=-e_{j}\wedge e_{i}=-e_{j}e_{i}.
\end{equation}%
We can also see that $ \iGA $ squares to minus one, that is $ \iGA^2 = e_{1}e_{2}e_{3} e_{1}e_{2}e_{3} = e_{1}e_{2} e_{1}e_{2} = -1 $
and commutes with all other elements and so has identical properties to the imaginary number~$ \jGA = \sqrt{-1} $. The ubiquitous presence of $ \sqrt{-1} $ in quantum mechanics, has been shown by Hestenes \cite{GA2} to be due to the presence of spin.  We thus give the unit imaginary `i' its appropriate geometrical representation as the trivector $ i = e_1 e_2 e_3 $.

The geometric product is in general not commutative but it is always associative, obeying $	a(bc) = (ab)c $.  
We also find the very useful result that the geometric product of a vector $ \textbf{v} $ with itself, yields the scalar dot product, that is $ \mathbf{v} \mathbf{v} = \mathbf{v} \cdot \mathbf{v} + \iGA \mathbf{v} \times \mathbf{v} = \mathbf{v} \cdot \mathbf{v}  = v_1^2 + v_2^2 + v_3^2 $.

Thus, from a pedagogical perspective, GA can more easily integrate with students' established intuition for elementary algebraic operations except that the concept of non-commutivity needs to be absorbed. However, it is actually a natural concept.  For example the subtraction operation is non-commutative!  That is $ 5-3 \ne 3 - 5 $.  A more physical example of non-commutivity is a series of rotations in three-dimensional space.

In order to progress to four dimensional spacetime, we now add $ e_0 $ as the time axis, which anti-commutes with the space coordinates $ e_k $ with $ e_0^2 = 1 $.  We then form a spacetime coordinate four-vector as
\be
\underline{x} = c t e_0 + x_1 e_1 e_0 + x_2 e_2 e_0 + x_3 e_3 e_0 = (c t+ \textbf{x}) e_0,
\ee
where we denote a four-vector using the underline.
This produces $ \underline{x}^2 = (c t+ \textbf{x}) e_0 (c t+ \textbf{x}) e_0 = (c t+ \textbf{x}) (c t- \textbf{x}) = c^2 t^2 - \textbf{x}^2 $, with the correct invariant interval in spacetime.  We could also define the product between four-vectors as $ \underline{x} \underline{x}^{\dagger} $ in order to produce the same result. A four-vector is typically written $ \underline{v} = [ v_0,v_1,v_2,v_3] = [ v_0,\textbf{v}] $, which is also the combination of a scalar with a vector, however in GA we provide the more general context for four-vectors as the two lowest orders of a multivector.  	

We can see that the correct sign structure is produced through $ e_0 $ anti-commuting with each of the space axes $ e_1,e_2,e_3 $.  That is we do not need to add a metric to the basis, retaining a consistent $ e_0^2 = e_1^2 = e_2^2 =e_3^2 =1 $ with all vectors anti-commuting equally with each other. Thus we have achieved Minkowski's original vision of creating a four-dimensional spacetime with a complete symmetry between the space and time coordinates \cite{EinsteinEtAl}.

So, summarizing our proposed notation for GA, following Hartle \cite{Hartle2003}$^,$ \cite{Rowland:2009}, we will use plain lower case letters for real scalars, (imaginary numbers are never used), plain upper case letters refer to general multivectors, unless indicated to be a vector by writing it in bold and with the unit three-vector written $ \mathbf{\hat{v}} $, four-vectors represented with an underline and the unit trivector given by $ \iGA $. A general multivector can be written
\be \label{generalMultivector}
M = a + \textbf{v} + \iGA \textbf{w} + \iGA b ,
\ee
which shows in sequence, scalar, vector, bivector and trivector terms and can be used to represent many mathematical objects, such as scalars ($a$), complex numbers ($ a + \iGA b $), quaternions ($a + \iGA \textbf{w}$), vectors (polar) ($\textbf{v}$), four-vectors ($a + \textbf{v}$), pseudovectors ($\iGA \textbf{w})$, anti-symmetric tensors ($\textbf{v} + \iGA \textbf{w}$) with its dual ($ \iGA \textbf{v} - \textbf{w}$) and spinors ($a + \iGA \textbf{w}$), with the four complex component Dirac spinor requiring the full multivector $ M $. Refer to Appendix D for explicit mappings to the GA multivector representation, demonstrating how GA unifies a diverse range of mathematical formalisms.
All products are the geometric product unless otherwise indicated.

We now proceed to describe Maxwell's equations in GA, followed by the potential formalism that naturally follows and offers further interesting insights.  We then describe the key equations used in solving problems in electromagnetism, such as the calculation of forces, and the conservation of energy and momentum.

The use of the Lorentz transformations are then described, as well as the modifications required to Maxwell's equation for the existence of magnetic monopoles.  

\section{Maxwell's equations in Geometric Algebra (GA)}

We can re-arrange Eq.~(\ref{eq:maxwellGA3DVold}) using the four-gradient $ \Box $ notation as
\be \label{eq:maxwellGA3DBoxFour}
\Box  F = c \mu_0 J,
\ee
where 
\bea \label{FJDefns}
\Box & = &   (\frac{1}{c} \partial_t - \mathbf{\nabla}) e_0 , \\ \nonumber
F & = & \mathbf{E}+\iGA c \mathbf{B}   =  E_x e_1 + E_y e_2 + E_z e_3  + \iGA c ( B_x e_1+ B_y e_2 + B_z e_3 ) ,  \\ \nonumber
J & = & (c \rho + \mathbf{J} ) e_0 ,    \\ \nonumber
\eea
where $ \iGA = e_1 e_2 e_3 $.  The $ \Box $ symbol represents the four-gradient, and we find $ \Box^2 = (\frac{1}{c} \partial_t - \mathbf{\nabla}) e_0 (\frac{1}{c} \partial_t - \mathbf{\nabla}) e_0  = (\frac{1}{c} \partial_t - \mathbf{\nabla}) (\frac{1}{c} \partial_t + \mathbf{\nabla})  = \frac{1}{c^2} \partial_t^2 - \nabla^2 $ giving the d'Alembertian.  It should be noted that there is a wide variation in notation within the literature when representing the d'Alembertian, such as $ \Delta, \Box, $ or $ \nabla $; however following Feynman \cite{Feynman1964}, we represent the d'Alembertian by $ \Box^2 $ and hence the four-gradient by $ \Box $.  

For a moving charge distribution creating a current we have  $ J =  \rho \underline{v}  $, where the four-velocity $ \underline{v} = \gamma ( c + v_x e_1  + v_y e_2  + v_z e_3 ) e_0  $ and we find $ \underline{v}^2 = c^2 $ as expected. For free magnetic monopoles $ \rho^m $, we have the additional bivector and trivector current terms given by $ \iGA \rho^m \underline{v} $. The source current $ J $, as defined in Eq.~(\ref{FJDefns}), is a four-vector, however we will not represent this with an underline in this case,  because as was just demonstrated, with the inclusion of magnetic monopoles, $ J $ becomes a full multivector.  With the addition of monopoles, the four-potential $ A $ also becomes a full multivector. From the four-velocity, we have the four-momentum defined as $ \underline{p} = m \underline{v} $, giving $ \underline{p}^2 = m^2 c^2 $. 

We also find $ \Box \cdot J =  (\frac{1}{c} \partial_t - \mathbf{\nabla}) e_0 \cdot (c \rho + \mathbf{J} ) e_0 = \frac{ \partial \rho }{\partial t } + \mathbf{\nabla} \cdot \textbf{J} $, so that setting the four-divergence to zero implies conservation of a quantity.

Eq.~(\ref{eq:maxwellGA3DBoxFour}) has been shown \cite{Vold:1993} using Green's functions to have a solution
\be \label{eq:FieldSolutionIntegral}
F(\mathbf{r},t) = \left (\frac{1}{c} \partial_t - \mathbf{\nabla} \right ) \frac{\mu_0}{4 \pi}  \int_{\rm{Vol}} \frac{c  \rho -\mathbf{J}}{r'} d \tau' = \Box \frac{\mu_0}{4 \pi} \int_{\rm{Vol}} \frac{J'}{r'} d \tau' ,
\ee
where $ r' = | \mathbf{r} - \mathbf{r'} | $, the distance from the field point $ \mathbf{r} $ to the charge at $ \mathbf{r'} $. The prime denoting the value at the retarded time $ t_r $.  Hence we can define a potential function 
\be \label{eq:VectorPotentialIntegral}
A= \frac{\mu_0}{4 \pi} \int_{\rm{Vol}} \frac{J'}{r'} d \tau' ,
\ee
giving $ F = \Box A $.

\subsection{Potential formulation in GA}

When moving to the potential formulation, in order to simplify the equations, the Lorenz gauge is typically introduced (sometimes erroneously called the Lorentz gauge \footnote{The Lorenz gauge is named after Ludvig Lorenz, whereas Lorentz covariance is with respect to Hendrik Lorentz.}).
Heras et al \cite{Heras:2010} show that the Lorenz gauge is the only gauge that puts both the electromagnetic fields and the potentials on a causal basis with the sources.  Also by multiplying both terms by $ q c $, they take on the dimensions of energy, thus it can be seen that the Lorenz gauge expresses the conservation of energy.

Looking at the Lorenz gauge in GA, we find it has the simple form of $ \Box \cdot A = c \mathbf{\nabla} \cdot \mathbf{A} + \frac{1}{c} \frac{\partial V}{\partial t} =  0 $, as shown in Table~\ref{tableFormalisms}.  Hence the Lorenz gauge written in GA is equivalent to setting the scalar part of the geometric product to zero and to relax this restriction we simply use the full geometric product, so that we then have simply 
\bea \label{eq:FWithoutMonopoles}
F & = & \Box A \\ \nonumber
& = & \left (\frac{1}{c} \partial_t - \mathbf{\nabla} \right ) e_0 \left (V  + c \mathbf{A} \right ) e_0  \\ \nonumber
& = & \left (\frac{1}{c} \partial_t - \mathbf{\nabla} \right ) \left (V  - c \mathbf{A} \right )  \\ \nonumber
& = & c \left ( \mathbf{\nabla} \cdot \mathbf{A}+\frac{1}{c^2} \partial_t V \right )+  c \mathbf{\nabla} \wedge \mathbf{A} - \mathbf{\nabla} V  - \partial_t \mathbf{A}   \\ \nonumber
& = & c \lambda - \mathbf{\nabla} V - \partial_t \mathbf{A}  + \iGA c \mathbf{\nabla} \times \mathbf{A} \\ \nonumber
& = & c \lambda + \mathbf{E} + \iGA c \mathbf{B} , \\ \nonumber
\eea
where the Lorenz gauge $ \lambda = \mathbf{\nabla} \cdot \mathbf{A} + \frac{1}{c^2} \frac{ \partial V}{\partial t} $.  Typically we set $ \lambda = 0 $, however the presence of a non-zero Lorenz gauge could now be further investigated.
Substituting into Eq.~(\ref{eq:maxwellGA3DBoxFour}) we produce the second order D.E.,
\be  \label{eq:maxwellGAPotentialForm}
\Box^2 A =  c \mu_0 J ,
\ee
where the four-vector $ A =    [V +c ( A_x e_1 + A_y e_2 + A_z e_3)] e_0=    (V +c \textbf{A})e_0 $ .
This naturally splits into four copies of Poisson's equation
\bea \label{eq:fourPoissonEquations}
\left (\frac{1}{c^2} \partial_t^2 - \mathbf{\nabla}^2 \right ) V & = & \frac{\rho}{ \epsilon_0 } ,  \\ \nonumber
\left (\frac{1}{c^2} \partial_t^2 - \mathbf{\nabla}^2 \right ) \mathbf{A} & = &  \mu_0 \mathbf{J} , \\ \nonumber
\eea
which have known solutions, as shown in Eq.~(\ref{eq:VectorPotentialIntegral}).  

The wave equation is generated from the case with no sources 
\be
\Box^2 A =  0 ,
\ee
with each component satisfying the three dimensional wave equation $ (\frac{1}{c^2} \partial_t^2 - \mathbf{\nabla}^2) \phi = 0 $.  The general solution of which is
\be \label{eq:WaveEquationGeneralSolution}
\phi(r,t) = \frac{1}{r} \big (F(r-ct) + G(r+ct) \big ),
\ee
where $ r =\sqrt{x^2+y^2+z^2} $.   The wave equation, including the polarization of light, can be further investigated using GA, as shown in \cite{Baylis:1992}. If we postulate that the photon is not completely massless, then we need to modify Eq.~(\ref{eq:maxwellGAPotentialForm}) as shown in Appendix E.

\subsection{Forces in the fields}

The force on a charge moving in an electromagnetic field is given classically by the Lorentz force law, as shown in Eq.~(\ref{LorentzForceLaw}).  We can write this in GA for a charge $ \rho $ moving with a four-velocity $ \underline{v} $, in a field $ F $ as
\bea \label{JFExpansion}
J  F & = & \rho \underline{v}  F  \\ \nonumber
& = &  \rho \gamma \left (c + \mathbf{v} \right ) e_0  \left (\mathbf{E} + \iGA \mathbf{B} \right ) =  \rho \gamma \left (c + \mathbf{v} \right ) \left (-\mathbf{E} + \iGA \mathbf{B} \right ) e_0  \\ \nonumber
& = & \gamma \left [-\rho \mathbf{v} \cdot \mathbf{E} - c \rho \left (\mathbf{E} - \iGA \mathbf{v} \wedge \mathbf{B} \right ) + \rho \mathbf{v} \wedge \textbf{E} + \iGA \mathbf{v} \cdot \textbf{B} \right ] e_0 \\ \nonumber
& = &  - \gamma \left [\rho \mathbf{v} \cdot \mathbf{E} + c \rho \left (\mathbf{E} + \mathbf{v} \times \mathbf{B} \right )  + \rho \mathbf{v} \wedge \textbf{E} + \iGA \mathbf{v} \cdot \textbf{B} \right ] e_0 , \\ \nonumber
\eea
which we can see consists of scalar, vector, bivector and trivector terms inside the square bracket.
The first term $ \rho \mathbf{v} \cdot \mathbf{E} $ is representing the power density and the second term is the Lorentz force, hence we have produced the Minkowski four-force $ \underline{K} $, so that we can write
\be \label{MinkowskiForce}
c \underline{K} =  - J F ,
\ee
where we simply disregard higher order terms from Eq.~(\ref{JFExpansion}) in order to produce the four-force~$ \underline{K} $.

\subsection{Conservation of energy - The work-energy theorem}

A key conceptual difference between Newtonian mechanics and electromagnetism is that instead of forces and energy conservation being connected to a localized object, these concepts now apply over a spread out field, and hence to apply the conservation of energy and momentum we now need to calculate a volume integral over the whole field, so that the total energy and momentum of both fields and matter is conserved.

The work-energy theorem of electromagnetism, also called Poyntings' theorem, is normally written \cite{Griffiths:1999} as
\be
- \mathbf{J} \cdot \mathbf{E} =   \frac{\partial u}{ \partial t}  +  \mathbf{\nabla} \cdot \mathbf{S} ,
\ee
where $ u $ the scalar potential energy and $ \mathbf{S} = \frac{1}{\mu_0}  \mathbf{E} \times \mathbf{B} = - \frac{\iGA}{\mu_0}  \mathbf{E} \wedge \mathbf{B}$ the Poynting vector, with each term having the units of power density.  However, we can write this using the four-gradient as $  \Box \cdot (c u + \mathbf{S}) e_0 = e_0 (\frac{1}{c} \partial_t + \mathbf{\nabla} ) \cdot (c u + \mathbf{S}) e_0  =   \frac{\partial u}{ \partial t}  +  \mathbf{\nabla} \cdot \mathbf{S} $, for the RHS, but also
\bea \label{FeoFExpansion}
\frac{c \epsilon_0}{2} F e_0 F & = & \frac{c \epsilon_0}{2} (\mathbf{E} + \iGA c \mathbf{B}) e_0 (\mathbf{E} + \iGA c \mathbf{B}) \\ \nonumber
& = & \big [- \frac{c \epsilon_0}{2} (\mathbf{E}^2 + c^2 \mathbf{B}^2 ) + \frac{c^2 \iGA \epsilon_0}{2} e_0 (\mathbf{E} \mathbf{B}-\mathbf{B} \mathbf{E} ) \big ]e_0 \\ \nonumber
& = & \big [- \frac{c }{2} ( \epsilon_0 \mathbf{E}^2 + \frac{1}{\mu_0} \mathbf{B}^2 ) - \frac{1}{\mu_0}  \mathbf{E} \times \mathbf{B} \big ] e_0  \\ \nonumber
& = & - ( c u +\mathbf{S}) e_0  \\ \nonumber
\eea
and hence we can write Poynting's theorem as
\be \label{workEnergyTheorem}
\mathbf{J} \cdot \mathbf{E}  =  c \, \Box \cdot T_0 ,
\ee
where
\be \label{StressEnergyTensorZero}
T_0 = \frac{1}{2} \epsilon_0 F e_0 F .
\ee
Also, we saw from the previous section that $ \textbf{J} \cdot \textbf{E} $ is the scalar component of $ J F $, therefore we can write Poynting's theorem in terms of $ F $ as
\be \label{workEnergyTheorem}
c \, \Box \cdot T_0 = J F .
\ee
Therefore energy conservation for the free fields can be written
\be \label{EnergyConservationFree}
\Box \cdot T_0 = 0 .
\ee
This is exactly parallel to charge $ \rho $ conservation with $ J = c \rho + \mathbf{J} $ given by $ \Box \cdot J = 0 $. Hence with $ c T_0 = c u + \textbf{S} $, we have energy $ u $ conservation given by $ \Box \cdot T_0 =  \frac{\partial u}{ \partial t}  +  \mathbf{\nabla} \cdot \mathbf{S} = 0 $. That is, if the energy $ u $ in the field falls, then this must be propagated away using the Poynting vector $ \mathbf{S} $.

\begin{figure}[htb]

\begin{center}
\includegraphics[width=4.5in]{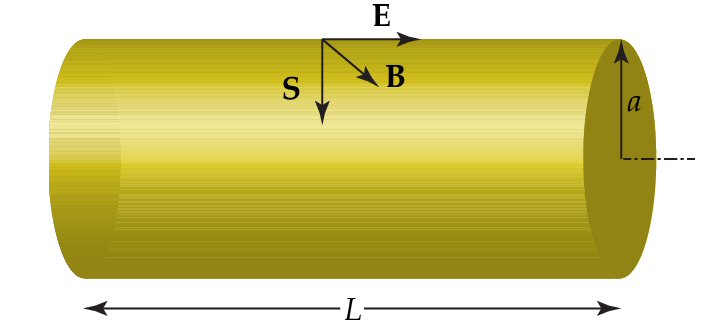}
\end{center}

\caption{Variables defined for calculation of energy dissipated in a wire.\label{powerWire}}
\end{figure}

For example for a circular wire, shown in Fig.~\ref{powerWire}, of radius $ a $ carrying a current $ I $, caused by a potential of $ V $ volts over its length $ L $, we have $ \textbf{E} = \frac{V}{L} e_z $ and $ \textbf{B}= \frac{\mu_0 I r }{2 \pi a^2} e_{\phi} $.

We find in cylindrical coordinates $ F = \frac{V}{L} e_z + \iGA \frac{c \mu_0 I r}{2 \pi a^2} e_{\phi} $ and hence 
\be
 T_0 = \frac{1}{2} \epsilon_0 \Big (\frac{V^2}{L^2} + \frac{c^2 \mu_0^2 I^2 r^2}{4 \pi^2 a^4 } \Big ) e_0 + \frac{ V I r}{2 c \pi a^2 L } e_r e_0.
\ee
We notice that $ T $ has no time dependence hence we only require the divergence $ \mathbf{\nabla} \cdot T_0 $ and so the only term that will contribute the required scalar is 
\be \label{WireTerm}
 \frac{ V I r}{2 c \pi a^2 L } e_r ,
\ee
which is the Poynting vector.
If we calculate the divergence and sum over the volume we firstly find $ \mathbf{\nabla} \cdot (r e_r) = 2 $ in cylindrical coordinates.
Hence the power per unit volume is $ c \, \Box \cdot T_0 = \frac{ V I }{ \pi a^2 L } $ and hence the total power for the volume $ \pi a^2 L $ of the wire is $ V I $, as expected.
Instead of integrating the divergence over the volume we can use the divergence theorem, and sum over the bounding area. On the surface of the wire we have $ r = a $ and hence multiplying Eq.~(\ref{WireTerm}) by the the surface area $ A = 2 \pi a L $ we again find the power dissipated as $ V I $.
As a third alternative from Eq. (\ref{workEnergyTheorem}), the power per unit volume is given by $  \mathbf{E} \cdot \mathbf{J} = \frac{V}{L} \frac{I}{\pi a^2 } $, and hence the total power over the volume of the wire is also $ V I $.

The conservation of momentum using the electromagnetic stress-energy tensor is introduced in Appendix H.

\subsection{Conservation of energy using the four-potential}

It is stated in \cite{Mead2002}, that the essence of electromagnetism, can be expressed in the interaction energy $ J \cdot A $, where $ A $ is the four-vector potential.
For example, for a loop of superconducting wire carrying a current, we have $ V = 0 $ and so ignoring the potential energy the total energy is kinetic
\bea
W_{\rm{kin}} & = & \frac{1}{2} \int_{\rm{vol}} \mathbf{A} \cdot \mathbf{J} d \tau \\ \nonumber
 & = & \frac{1}{2} \int_{\rm{vol}} J da \mathbf{A} \cdot \mathbf{dl}  \\ \nonumber
 & = & \frac{1}{2} I \int_{\rm{vol}} \mathbf{A} \cdot \mathbf{dl}  \\ \nonumber
 & = & \frac{1}{2} I \Phi  \\ \nonumber
 & = & \frac{L I^2}{2} ,  \\ \nonumber
\eea
using $ \Phi = L I $, yielding the correct energy for an inductor.

\subsection{Lorentz transformations}

Simple experiments can confirm that moving a magnet past a wire will create an electric current or conversely a changing electric field will create a magnetic field.  This can be viewed quite generally as a result of placing the observer into a frame of relative motion.  

When an observer moves past an electric field, it is found that the field parallel to the relative velocity is unaffected. Hence it is natural to split the field into two components parallel and perpendicular to the direction of motion, that is $ \mathbf{E} = \mathbf{E}_{\|} + \mathbf{E}_{\perp} $.  If the relative four-vector motion $ \underline{v} $ between the frames of reference, is represented by a unit four-vector $ u = \frac{1}{c} \underline{v} e_0= \gamma ( 1 + \frac{\mathbf{v}}{c} ) $, then the transformed fields are found simply by appending this relative four-velocity onto the perpendicular components of the field, that is
\be \label{FieldsTransform}
F' = \mathbf{E}_{\|} + \mathbf{E}_{\perp} u .
\ee
For example, if we wish to calculate the electric and magnetic fields of a charge $ Q $ moving at a speed $ v/c $ in the $ e_1 $ direction relative to our instruments, then we can firstly look at the field from the simpler perspective of the frame at rest with respect to the charge. So, using Coulomb's law we have simply
\be
\mathbf{E} = \frac{1}{4 \pi \epsilon_0} \frac{Q}{r^2} \mathbf{\hat{r}} .
\ee
We then transform this into the moving observer frame, using Eq.~(\ref{FieldsTransform}), giving
\bea
F' & = & E_x e_1  + (E_y e_2 + E_z e_3 ) \gamma ( 1 + \frac{v}{c} e_1 ) \\ \nonumber
 & = & E_x e_1  + \gamma (E_y e_2 + E_z e_3 ) + \frac{v}{c} \gamma (E_y e_2 + E_z e_3 ) e_1   \\ \nonumber
 & = & E_x e_1  + \gamma (E_y e_2 + E_z e_3 ) + \frac{v}{c} \gamma (-E_y e_1 e_2 + E_z e_3 e_1 )    . \\ \nonumber
\eea
Hence we have
\bea
E_x'=E_x  , \, E_y'  = \gamma E_y , \, E_z' = \gamma E_z \\ \nonumber
B_x' = 0   , \, c B_y'  = \gamma \frac{v}{c} E_z , \, c B_z' = - \gamma \frac{v}{c} E_y , \\ \nonumber
\eea
which are the correct transformation laws for a pure electric field \cite{Griffiths:1999}, demonstrating how magnetic fields appear through relative motion.

A similar approach works when translating coordinates of events between different observer frames, except that in this case we find that the perpendicular components are unaffected. Writing the four-coordinate $ \underline{x} = \underline{x}_{\perp}+ \underline{x}_{\| } $ we then find the correct transformation is found by appending the relative velocity $ u $ onto the parallel components of the coordinates, that is
\be \label{CoordinateBoost}
\underline{x}' = \underline{x}_{\|} u + \underline{x}_{\perp}.
\ee
For example for a boost in the $ e_1 $ direction of velocity $ v $ on coordinates $ \underline{x} = (ct + x e_1 + y e_2 + z e_3) e_0 $, we find using Eq.~(\ref{CoordinateBoost})
\bea
\underline{x}' & = & \left (ct + x e_1 \right ) e_0 \gamma \left (1+ \frac{v}{c} e_1 \right ) + \left (y e_2 + z e_3 \right ) e_0 \\ \nonumber
 & = & \left [ \left (ct + x e_1 \right ) \gamma \left (1 -  \frac{v}{c} e_1 \right ) + y e_2 + z e_3 \right  ] e_0 \\ \nonumber
 & = & \left [ \gamma \left (ct - x v/c \right ) + \gamma \left ( x  - vt \right ) e_1 + y e_2 + z e_3 \right ] e_0 , \\ \nonumber
\eea
which gives us the correct coordinate transformations
\bea
c t' & = & \gamma \left (c t - \frac{x v}{c} \right ), \,\,\, x' =  \gamma \left (x - v t \right ) , \,\,\, y' = y , \,\,\, z' = z . \\ \nonumber
\eea

More generally, if it is not convenient to extract parallel and perpendicular components from the field, then we can undertake the Lorentz transformations of the field or coordinates as a whole using a rotor. If we have a frame moving with velocity $ v = |\textbf{v}| $ in the $ \mathbf{\hat{v}} $ direction, then the field $ F $ transforms according to
\be
F' = R F R^{\dagger},
\ee
where $ R  = \rme^{- \theta \mathbf{\hat{v}} /2} $ representing a boost in the $ \mathbf{\hat{v}} $ three-vector direction, where $ \theta $ is given by $ \tanh \theta =  v/c $ and hence $ \gamma = \cosh \theta $. We have also defined $ R^{\dagger} = e_0 \tilde{R} e_0 $ that corresponds to the Hermitian conjugate, where $ \tilde{R} $ reverses the order of products.  The identical transformation law works to boost coordinates $ \underline{x} = (c t + \textbf{x}) e_0 $, being $ \underline{x}' = R \, \underline{x} R^{\dagger} $.  Thus the GA formalism has the advantage of using exactly the same transformation law for boosts, for both coordinates and fields.
We can also do spatial rotations using the rotor $ S = \rme^{i \mathbf{\hat{u}} \phi/2} $, when we are rotating about the axis of the unit vector $ \mathbf{\hat{u}} $.

\section{Summary of key relations in electromagnetism using GA}

We can now summarize the relations required in GA to solve a typical set of electromagnetic problems.  
Given sources we have the field equations
\be
\Box F = c \mu_0 J,
\ee
or in potential form
\be
\Box^2 A = c \mu_0 J
\ee
where $ F = \textbf{E} + \iGA c \textbf{B} = \Box A $.  With fields defined in linear isotropic dielectrics we have $ F = c \mu_0 ( c \textbf{D} + \iGA \textbf{H} ) $, as seen in Appendix F.

The conservation principles are
\bea
\Box  \cdot J & = & 0 , \,\,\,\, \rm{Conservation \, of \, charge} \\ \nonumber
\Box  \cdot T & = & 0 , \,\,\,\, \rm{Conservation \, of \, momentum \, and \, energy} \\ \nonumber
\Box  \cdot A & = & 0 , \,\,\,\, \rm{Conservation \, of \, energy \, (potential \, formulation). } \\ \nonumber
\eea
We have the Minkowski four-force 
\be
c \underline{K} = - q \underline{v} F .
\ee

Through the use of the relativity principle we can work in a co-moving frame and convert four vectors simply by piggybacking the parallel component of the co-ordinate onto the relative velocity vector $ u $
\be
\underline{x}' = \underline{x}_{\|} u + \underline{x}_{\perp}
\ee
or alternatively, both fields and four-vectors can be transformed using the Lorentz transformation
\be
a' = R a R^{\dagger}
\ee
where $ R = \rme^{- \theta \mathbf{\hat{u}} /2} $ and $ a $ can represent either a four-vector coordinate or the field $ F $.

\section{Extended topics}

We now present two specialized topics presented in the formalism of GA.  An introduction to the Lagrangian formulation of electromagnetism using GA is also placed in Appendix G.

\subsection{Magnetic monopoles in GA}

One extension of Maxwell's equations that is commonly considered is the inclusion of magnetic monopoles. As mentioned, no free monopoles have yet been found, however a recent result in 2009 demonstrated an effective flow of magnetic monopoles in a spin ice system \cite{monopoles:2009}, though they cannot exist outside the material in a free form. There is an ongoing search for free monopoles at high energies \cite{Cozzi2006} and so may yet be discovered. Hence the presence of monopoles should be allowed for by expanding Maxwell's equations.
Fortunately these can be included very simply by expanding out the RHS of Eq.~(\ref{eq:maxwellGA3DBoxFour}) to a full multivector, that is, by including the bivector and trivector terms in the source $ J $ as shown,
\be \label{eq:maxwellGASISimpleFinalMagneticMonopoles}
(\frac{1}{c} \partial_t + \mathbf{\nabla}) ( \mathbf{E} + \iGA c \mathbf{B}) =  c \mu_0 ( c \rho - \mathbf{J}  - \frac{\iGA}{c} \mathbf{J^m} + \iGA \rho^m  ) ,   
\ee
where $ \rho^m $ represents magnetic charge and $ J^m $ is the current of magnetic charge.
So we can see that we now have $ \nabla \cdot B =  \mu_0 \rho^m $ and $ \mathbf{\nabla} \times \mathbf{E} +  \partial_t \mathbf{B} = - \mu_0 \textbf{J}^m $ as required for the presence of monopoles.  

Thus the use of GA leads to obvious `holes' in Eq.~(\ref{eq:maxwellGA3DBoxFour}),  which can very naturally be filled with the terms for magnetic monopoles in the multivector representing the source currents.  Also we can see that we can generalize no further because the multivector for $ J $ is now complete.

Thus GA provides a natural framework for further investigation of the properties of monopoles.

\subsection{Lienard-Wiechert potentials}

For a single charge $ q $ in arbitrary motion with the four-vector distance between the charge and the field point $ \underline{r} = (ct + \mathbf{r} ) e_0 $, with four velocity $ \underline{v}_r = (c + \mathbf{v}_r) e_0 $ at the retarded time, we can write the four-vector potential as
\be
A(r,t) = \frac{1}{4 \pi \epsilon_0} \frac{q \underline{v}_r}{\underline{r} \cdot \underline{v}_r} ,
\ee
where to place the charge on the past light cone, we specify $ \underline{r}^2 = 0 $ or $ c^2 t^2 -\mathbf{r}^2 = 0 $ that implies $ |\mathbf{r}| = ct $.  Therefore we find $ \underline{r} \cdot \underline{v}_r = (ct + \mathbf{r} ) e_0 \cdot (c + \mathbf{v}_r) e_0 = c^2 t - \mathbf{r} \cdot \textbf{v}_r = c |r| - \mathbf{r} \cdot \mathbf{v}_r$. Thus
\be
A(r,t) = \frac{1}{4 \pi \epsilon_0} \frac{q (c + \mathbf{v}_r) e_0}{c |r| - \mathbf{r} \cdot \mathbf{v}_r} ,
\ee
producing the well known Lienard-Wiechert potentials \cite{Griffiths:1999}
\be
V(r,t) = \frac{1}{4 \pi \epsilon_0} \frac{q c}{ c |r| - \mathbf{r} \cdot \mathbf{v}_r} \,\,\, , \mathbf{A}(r,t) = \frac{\mathbf{v}_r}{c^2} V(r,t),
\ee
where $ \textbf{r} $ is the vector from the observer to the moving charge at the retarded time $ t_r $ given by $ | r | = | \mathbf{r} - \textbf{w}(t_r)| = c (t-t_r) $, where $ \textbf{w} $ is the position vector of the moving charge, moving at a velocity $ \mathbf{v}_r $.

We could now calculate the $ \mathbf{E} $ and $ \mathbf{B} $ fields by differentiating these potentials, alternatively we can simply calculate the Coulomb field in a co-moving frame, and apply a Lorentz transformation to find the correct $ \mathbf{E} $ and $ \mathbf{B} $ fields, find the electric field
\be
\mathbf{E} = k \frac{q}{4 \pi \epsilon_0 } \frac{\mathbf{\hat{r}}}{r^2} ,
\ee
where $ k $ is a scaling factor depending on the relative velocity and the angle of the velocity vector of the charge.

So, for a charge in uniform motion, the electric field vector $ \textbf{E} $ as seen by the observer, there is an interesting quirk of nature, in that the electric field vector points towards where the charge is now instantaneously \cite{Griffiths:1999}, indicated by $ \mathbf{\hat{r}} $, not where the charge was when it generated the field at the retarded time.  This is a significant simplification for this special class of problems where charges and observers are in uniform relative motion, giving a convenient shortcut for students.

\section{Discussion}

We have shown how the GA formalism provides a simple algebraic approach to solving several common situations encountered in the field of electromagnetism, while maintaining manifest Lorentz covariance in all our equations.
The formalism can achieve this without the significant overhead of the identities for vector calculus or the index notation of tensors and its associated spacetime metric.

We have also focused on a straightforward presentation for the calculation of forces and the equations for the conservation of energy and momentum, in order to allow an easy progression from experience gained in Newtonian mechanics.

The potential formulation also neatly integrates into the presentation, which provides an alternate solution path for some problems.  We adopt the Lorenz gauge that guarantees causality in all expressions. The Lorentz transformations take a particularly simple form in GA and allows easy use of the co-moving frame.

The relative simplicity of GA  is also illustrated visually by inspection of Table~\ref{tableFormalisms} and we would expect a much faster learning curve for students using the formalism of GA.
It also should be noted that all results and equations are naturally in a full relativistic setting, and so no re-learning is required when relativistic effects are included.

In higher dimensions than two, for example with spinors, it is typically difficult for students to visualize the complex numbers, but because in GA the imaginary $ \sqrt{-1} $ has been replaced with the trivector $ i = e_1 e_2 e_3 $, results can immediately plotted on the three-space axes $ e_1, e_2, e_3 $, allowing geometric intuition to be more easily developed.

In order to introduce new students to the properties of electric and magnetic fields, it has been found that a series of basic experiments on the field properties are best undertaken and the relevant Maxwell equation introduced in sequence.  Using GA, an overview can be given at each step, showing how each equation can be extracted from the master equation, Eq. (\ref{eq:maxwellGA3DVold}), thus providing students with an overview, or a mind map within which new results can be placed.  This would also be beneficial for past students to the field, who are familiar with the physical experiments and can now be given a unified mathematical framework.

\section{appendix}

\subsection{The equivalence of the GA form of Maxwell's equations}

From Eq.~(\ref{eq:maxwellGA3DVold}), Maxwell's equations in GA can be written out fully as
\bea \label{eq:maxwellGASISimpleFinalAppendix}
& & \left (\frac{\partial_t}{c}  + e_1 \partial_x+ e_2 \partial_y+ e_3 \partial_z \right ) \left [E_x e_1 + E_y e_2 + E_z e_3  + c \left ( B_x e_2 e_3+ B_y e_3 e_1 + B_z e_1 e_2 \right ) \right ] \\ \nonumber
& = & \frac{\rho}{ \epsilon_0 }   - c \mu_0 \left (J_x e_1 + J_y e_2 + J_z e_3 \right ). \\ \nonumber
\eea
This form has all dot products and cross products expanded in full, but formed into a single equation.
This form reproduces Maxwell's equations without the use of a metric using just the three Euclidean space directions $ e_1,e_2,e_3 $.  

Using the relations $ e_i e_j = \delta_{ij} $ and  the anti-commuting property $ e_i e_j = - e_j e_i  $, we can firstly equate scalar parts, to confirm that $ \mathbf{\nabla} \cdot \mathbf{E}  = \frac{\rho}{\epsilon} $.  Next, by collecting the $ e_1 $,$ e_2 $ and $ e_3 $ terms we find
\bea
  \frac{1}{c } \partial_t E_x e_1 - e_1 c (\partial_y B_z - \partial_z B_y)   & = &  -c \mu_0 \gamma_1 J_x  \\ \nonumber
  \frac{1}{c } \partial_t E_y e_2 - e_2 c ( -\partial_x B_z + c\partial_z B_x)   & = & - c \mu_0 \gamma_2 J_y  \\ \nonumber
  \frac{1}{c } \partial_t E_z e_3 - e_3 c (\partial_x B_y - c\partial_y B_x)   & = &  - c \mu_0 \gamma_3 J_z,  \\ \nonumber
\eea
which when added give $  \mathbf{\nabla} \times \mathbf{B} - \frac{1}{c^2 } \partial_t \mathbf{E} =   \mu_0 \mathbf{J} $ as required by Amp$\grave{\rm{e}}$re's law.
We have the $ \iGA $ trivector terms, giving
\be
c \iGA (\partial_x B_x+\partial_y B_y+\partial_z B_z) = 0 
\ee
giving $  \mathbf{\nabla} \cdot \mathbf{B}  = 0 $.  

The $  e_1 e_2 $, $   e_1 e_3  $ and $  e_2 e_3  $ terms give
\bea
 e_2 e_3 \partial_t B_x  + e_2 e_3 ( \partial_y E_z - \partial_z E_y)  & = & 0 \\ \nonumber
 e_3 e_1  \partial_t B_y  +   e_3 e_1   ( -\partial_x E_z + \partial_z E_x )  & = & 0 \\ \nonumber
  e_1 e_2 \partial_t B_z  +  e_1 e_2  ( \partial_x E_y - \partial_y E_x )  & = & 0  \\ \nonumber
\eea
and adding these three equations we find $ \mathbf{\nabla} \times \mathbf{E}+ \partial_t \mathbf{B} = 0 $,
thus Eq.~(\ref{eq:maxwellGA3DVold}) reproduces all four of Maxwell's equations.

\subsection{Introduction to tensor formalism}

Using tensors we can bring Maxwell's four equations down to two.
We construct an anti-symmetric field tensor
\be
F^{\mu \nu } = \begin{bmatrix} 0 & E_x & E_y & E_z \\
-E_x & 0 & c B_z & - c B_y\\
-E_y & -c B_z & 0 & c B_x \\
-E_z & c B_y & -c B_x & 0  \end{bmatrix} 
\ee
with the dual tensor $ G^{\mu \nu } = \epsilon^{\alpha \beta \mu \nu} F_{\alpha \beta}  $ and if we define the current density 
\be
J_{\mu} = \rho_0 v^{\mu} = \gamma \rho_0 (c,v_x,v_y,v_z) = ( c \rho,J_x,J_y,J_z) = ( c \rho,\textbf{J}),
\ee
then Maxwell's equations read
\be
F^{\alpha \beta }_{, \alpha} = c \mu_0 J^{\beta}  , \,\,\, \epsilon^{\alpha \beta \gamma \delta} F_{\alpha \beta , \gamma} = 0 ,
\ee
where the comma in front of a symbol implies differentiation with respect to this coordinate.
That is we can define
\be
F^{\mu \nu} =  \frac{\partial A^{\nu}}{\partial x^{\mu}} - \frac{\partial A^{\mu}}{\partial x^{\nu}} = A^{\nu}_{,\mu} - A^{\mu}_{,\nu}
\ee
and with the Lorenz gauge $ A^{\mu}_{, \mu} = 0 $ we have
\be
\Box^2 A^{\mu} = - c \mu_0 J^{\mu} ,
\ee
where the 4-vector potential $ A^{\mu} = ( V,c A_x,c A_y,c A_z) $.

If we have the proper velocity $ v_{\nu} = \frac{d x^{\mu}}{d \tau} = \gamma ( c, v_x,v_y,v_z) $, where $ v $ is the velocity relative to an observer and $ \gamma = \frac{1}{\sqrt{1-v^2/c^2}} $ then the 4-momentum is defined as $ \underline{p} = m \underline{v} $. The Minkowski force on a charge $ q $
\be
K^{\mu} = \frac{d p^{\mu}}{d \tau} = \frac{d t}{d \tau} \frac{d p}{d t} = \gamma F = q v_{\nu} F^{\mu \nu} .
\ee
Vectors transform as $ \tilde{a}^{\mu} = \Lambda^{\mu}_{\nu} a^{\nu} $ and for the field tensor $ \tilde{F^{\mu \nu}} = \Lambda^{\mu}_{\lambda} \Lambda^{\nu}_{\sigma} F^{\lambda \sigma} $.

We have the spacetime metric is given by
\be
g^{\mu \nu } = \begin{bmatrix} 1 & 0 & 0 & 0 \\
0 & -1 & 0 & 0\\
0 & 0 & -1 & 0 \\
0 & 0 & 0 & -1  \end{bmatrix} .
\ee

\subsection{Calculating the outer product} 

The question often arises, of exactly how do I calculate the outer product?
There are two ways to calculate the outer product.  Firstly, if we are in three dimensions, then we can use the relation that $ \textbf{u} \wedge \textbf{v} = \iGA \textbf{u} \times \textbf{v} $.
Using the calculation method typically taught for the cross-product, using the determinant of the matrix
\be
\begin{bmatrix} e_1 & e_2 & e_3 \\
u_1 & u_2 & u_3 \\
v_1 & v_2 & v_3  \end{bmatrix} 
\ee
we find
\be
\textbf{u} \times \textbf{v} = (u_2 v_3 - u_3 v_2 ) e_1 - (u_1 v_3 - u_3 v_1 ) e_2 + (u_1 v_2 - v_1 u_2 ) e_3 .
\ee
Hence, using $ \iGA = e_1 e_2 e_3 $, we find
\bea
\textbf{u} \wedge \textbf{v} & = & \iGA \textbf{u} \times \textbf{v} \\ \nonumber
& = & e_1 e_2 e_3 [(u_2 v_3 - v_2 u_3 ) e_1 - (u_1 v_3 - u_3 v_1 ) e_2 + (u_1 v_2 - v_1 u_2 ) e_3] \\ \nonumber
& = & (u_2 v_3 - v_2 u_3 ) e_2 e_3  + (u_1 v_3 - u_3 v_1 ) e_1 e_3 + (u_1 v_2 - v_1 u_2 ) e_1 e_2 . \\ \nonumber
\eea
More generally, we can find the outer product simply by expanding the components of the vectors as a geometric product and then deleting the scalar part, that is we find firstly for the geometric product
\bea
\textbf{u} \textbf{v} & = & u_1 v_1 + u_2 v_2 + u_3 v_3 \\ \nonumber
 & & + (u_2 v_3 - v_2 u_3 ) e_2 e_3 + (u_1 v_3 - u_3 v_1 ) e_1 e_3 + (u_1 v_2 - v_1 u_2 ) e_1 e_2, \\ \nonumber
\eea
using $ e_i^2 = 1 $ and $ e_i e_j = - e_j e_i $, see Eq.~(\ref{VectorProductExpand}).
We know $ \textbf{u} \textbf{v} = \textbf{u} \cdot \textbf{v} + \textbf{u} \wedge \textbf{v} $ and that the dot product is a scalar, hence we simply remove the scalar part of the geometric product to give the outer product
\be
\textbf{u} \wedge \textbf{v}  = (u_2 v_3 - v_2 u_3 ) e_2 e_3 + (u_1 v_3 - u_3 v_1 ) e_1 e_3 + (u_1 v_2 - v_1 u_2 ) e_1 e_2
\ee
in agreement with result calculated using the cross product.  However this approach, of expanding the brackets, works for vectors in any number of dimensions and so is more general as well as being more intuitive.

Also, using the fact that the dot product for vectors is commuting and the outer product is anti-commuting we can write the specific relations
\bea
\textbf{a} \cdot \textbf{b} = \frac{1}{2} ( \textbf{a} \textbf{b} + \textbf{b} \textbf{a} ) \\ \nonumber
\textbf{a} \wedge \textbf{b} = \frac{1}{2} ( \textbf{a} \textbf{b} - \textbf{b} \textbf{a} ), \\ \nonumber
\eea
which enables us calculate these two products directly in terms of geometric products.

Not just vectors, but general multivector products can be evaluated using the distributive law to expand the brackets and the anti-commutitive property of the orthonormal basis vectors $ e_i e_j = - e_j e_i $ and the orthonormal property $ e_i^2 = 1 $.  As an example, the following product can be expanded as follows
\bea
& & ( e_3 + 2 e_2 e_3 + i)(2 + 5 e_1 e_2) \\ \nonumber
& = &  2 e_3 + 4 e_2 e_3 + 2 i + 5 i - 10 e_1 e_3 - 5 e_3 \\ \nonumber
& = & -3 e_3 + 4 e_2 e_3  - 10 e_1 e_3  + 7 i .\\ \nonumber
\eea
We can see how irreducible elements of the algebra ($1,e_i, e_i e_j, \iGA $), transform into each other and can be freely combined together when using GA.
For example when solving Eq.~(\ref{eq:maxwellGA3DVold}), because the LHS, when expanded using the geometric product, can contain all eight possible algebraic elements, $1,e_i, e_i e_j, \iGA $, then this single equation actually represents eight separate equations, which for a the field vector $ F = \textbf{E} + \iGA \textbf{B} $, will have six unknowns as the scalar and trivector terms are missing.  This is similar to equating real and imaginary parts in a complex expression, except that in this case we have eight distinct elements to equate.

\subsection{Mapping to the multivector}

The general three-space multivector in GA, shown in Eq.~(\ref{generalMultivector}), can be used to represent a diverse number of mathematical objects.
Firstly, we can obviously represent scalars ($a$), vectors ($\textbf{v}$) and complex numbers ($a+ \iGA b$), where $ i = e_1 e_2 e_3 $.  However the quaternions  i,j,k, defined by $ \rm{i}^2 = \rm{j}^2 = \rm{k}^2 = \rm{i j k} = -1 $, maps into the multivector $\iGA \textbf{w} $, or component wise, $ {\rm{i}} \rightarrow \iGA e_1 , {\rm{j}} \rightarrow  - \iGA e_2 , {\rm{k}} \rightarrow \iGA e_3 $.  Hence a general quaternion
\be
q = a + b {\rm{i}}  - c {\rm{j}} + d {\rm{k}} \leftrightarrow  a + \iGA b e_1 + \iGA c e_2 + \iGA d e_3 = a + \iGA \textbf{w} .
\ee 
This is similar to the Pauli spinor mapping 
\be \label{SpinorMapping}
| \psi \rangle = \begin{bmatrix} a + j a_3  \\ -a_2 + j a_1  \end{bmatrix} \leftrightarrow \psi = a + i a_1 e_1 + i a_2 e_2 + i a_3 e_3 = a + \iGA \textbf{w},
\ee
also mapping to the even subalgebra of the multivector, which shows the close relationship between spinors and quaternions.  A single qubit can also be represented as spinor in GA \cite{CIL}.

The electromagnetic field represented by the antisymmetric tensor $ F^{\mu \nu} $, see Appendix B, maps as follows
\be \label{FieldTensorMapping}
F^{\mu \nu} \leftrightarrow F = \textbf{E}+\iGA c \textbf{B} ,
\ee
with the dual tensor $ G^{\mu \nu} $ given in GA by $ G = \iGA F $. We also have other dual spaces in the multivector, such as for vectors $ \textbf{v} \leftrightarrow \iGA \textbf{w} $ and for spinors $ a+ \iGA \textbf{w} \leftrightarrow \iGA (b - \iGA \textbf{v}) $.

The Dirac wave function maps to the full multivector as follows
\be \label{DiracWaveFnMapping}
 | \psi \rangle = \begin{bmatrix} a + j a_3  \\ a_2 + j a_1  \\ - b_3 + j b  \\ -b_1 + j b_2  \end{bmatrix} \leftrightarrow  a_0 + b_1 e_1 + b_2 e_2 + b_3 e_3 + i a_1 e_1 + i a_2 e_2 + i a_3 e_3   + \iGA b = a + \textbf{v} + \iGA \textbf{w} + \iGA b  .
\ee
Other entities can also be mapped to a multivector, such as pseudoscalars ($\iGA b $) and pseudovectors ($ \iGA \textbf{w}$).
Rotation matrices are also represented in a multivector as $ R = a + \iGA \textbf{w} $, which rotates a vector about the $ \textbf{w} $ axis by an angle given by $ \tan \theta = |w|/a $ according to
\be
\textbf{v} \, ' = R \textbf{v} R^{\dagger} .
\ee
Hence the great capability of the three-space multivector is demonstrated, being able to replace a large variety of mathematical structures and formalisms as shown and hence unify the mathematical language of physics \cite{GA2}. 

\subsection{Proca Equation in GA}

Maxwell's equations implicitly assume that the photon has zero mass and current precision tests have concluded that its mass is indeed approaching zero, with an upper bound of $ m_{\gamma} \le 6 \times 10^{-17} $eV \cite{Tu2005}.  However even a very small mass would be significant in that it would imply an additional longitudinal polarization direction for light as well as implying a variation from the inverse square law.

Following Proca \cite{Gondran2009}, in order to correct Maxwell's equations, we add a mass term to Eq.~(\ref{eq:maxwellGAPotentialForm}) as follows 
\be \label{ProcaPotential}
\Box^2 A =  c \mu_0 J - \frac{m_{\gamma}^2 c^2}{\hbar^2} A ,
\ee
which in free space, $ J = 0 $, produces the Klein-Gordon equation.
The photon's mass can be assumed to be a real mass, or an effective mass which can be imparted in plasmas.

\subsection{Quasi-static expansion of Maxwell's equations using GA}

The quasi-static expansion of Maxwell's equations was used by Chua in order justify the existence of a new circuit element he called the memristor \cite{Chua1971}.  In the presence of linear isotropic dielectrics, Eq.~(\ref{eq:maxwellGA3DVold}) can be written
\begin{equation} \label{MaxwellDielectrics}
\left (\frac{1}{c} \partial_t + \mathbf{\nabla} \right ) F' = J
\end{equation}
where $ F' = c \textbf{D} + \iGA \textbf{H} $ and $J = c \rho - \textbf{J} $, where $ \textbf{D} = \epsilon \textbf{E} $ and $ \textbf{B} = \mu \textbf{H}$.
We can define the vector fields, as a power series in $\alpha $, for example
using the electric field we have 
\begin{eqnarray}
\textbf{E} &=&\textbf{E}_{\alpha =0}+\alpha \frac{\partial \textbf{E}}{%
\partial \alpha }\Big|_{\alpha =0}+\frac{\alpha ^{2}}{2}\frac{\partial ^{2}%
\textbf{E}}{\partial \alpha ^{2}}\Big|_{\alpha =0}+\dots +\frac{\alpha ^{k}}{%
k!}\frac{\partial ^{k}\textbf{E}}{\partial \alpha ^{k}}\Big|_{\alpha
=0}+\dots  \notag  \label{EPowerSeries} \\
&=&\textbf{E}_{0}+\alpha \textbf{E}_{1}+\alpha ^{2}\textbf{E}_{2}+\dots
+\alpha ^{k}\textbf{E}_{k}+\dots  \label{PowerSeriesE}
\end{eqnarray}%
where 
\begin{equation}
\textbf{E}_{k}=\frac{1}{k!}\frac{\partial ^{k}\textbf{E}}{\partial \alpha
^{k}}\Big|_{\alpha =0} .
\end{equation}%
We can then write
\begin{equation}
\left (\frac{\alpha}{c} \partial_{\tau} + \mathbf{\nabla} \right ) \left  (F_0 + \alpha F_1 +
\alpha^2 F_2 + \dots \right ) = \left (J_0 + \alpha J_1 + \alpha^2 J_2 + \dots \right ),
\end{equation}
hence it is easy to see that the orders of the quasistatic expansion become 
\begin{eqnarray}
\mathbf{\nabla} F_0 & = & J_0 \\
\partial_{\tau} F_0 + \mathbf{\nabla} F_1 & = & J_1  \notag \\
\partial_{\tau} F_1 + \mathbf{\nabla} F_2 & = & J_2  \notag \\
\dots & = & \dots  \notag \\
\notag
\end{eqnarray}
The process of solution is now very clear in GA. From the zeroeth order fields we
calculate $F_0 $, which is substituted into the first order fields to find $ F _1 $, and so on.

\subsection{Lagrangian formulation}

As an alternative to the Lorentz force law, we can derive the equations of motion from a Lagrangian, where $ L = T - U $, with $ T $ representing the kinetic energy, and $ U $ the potential energy.  The advantage of the Lagrangian approach is that just two scalar fields $ T $ and $ V $ are required, as opposed to the two vector fields, $ \textbf{E} $ and $ \textbf{B} $ as part of the Lorentz force law, shown in Eq.~(\ref{LorentzForceLaw}).
The Lagrangian for electromagnetism can be written in GA
\be
L = L_{\rm{field}}+L_{\rm{int}} =  \frac{1}{2 \mu_0} F^2 -  J A ,
\ee
with the interaction energy $ J \cdot A = q V - q \textbf{v} \cdot \textbf{A} $.
Substituting $ F = \Box A $ we find
\be \label{eq:LagrangianInA}
L =  \frac{1}{2 \mu_0} (\Box A)^2 -  J A.
\ee
Lagrange's equations for a field in tensor notation are
\be
\partial_{\nu} \Big ( \frac{\partial L}{\partial ( \partial_{\nu} A_{\mu})}\Big ) - \frac{\partial L}{\partial A_{\mu}} = Q_{\mu},
\ee
where $ Q_{\mu} $ refers to external forces.  Thus in four-space we form four equations of motion, one for each coordinate. In GA, Lagrange's equations become 
\be \label{eq:LagrangesEquationsInA}
\Box \Big ( \frac{\partial L}{\partial ( \Box A)}\Big ) - \frac{\partial L}{\partial A} = Q.
\ee
For a particle free falling in a potential field we have $ Q = 0 $, thus substituting the Lagrangian Eq.~(\ref{eq:LagrangianInA}) into Lagrange's equations Eq.~(\ref{eq:LagrangesEquationsInA}), we find
\be
\Box \Big (\frac{1}{\mu_0} \Box A \Big ) - J = 0 ,
\ee
and hence we recover
\be
\Box^2  A = \mu_0 J ,
\ee
very simply confirming the Lagrangian approach in GA.

The Lagrangian of a relativistic charged particle of mass $ m $, charge $ q $ and four velocity $ \underline{v} $, moving in an electromagnetic field with vector potential $ A $ we add a kinetic term to find
\be
L =  \frac{1}{2 \mu_0} F^2 - J A - \frac{1}{\gamma} m c^2 .
\ee

\subsection{Stress-energy tensor}

Momentum conservation in the tensor formalism is normally written
\be \label{MomentumConservationTensor}
\partial_{\nu} T^{\mu \nu} = F^{\alpha \mu} J_{\alpha} /c = K^{\mu}, 
\ee
where $ T^{\mu \nu} = \epsilon_0 ( F^{\mu \alpha } \eta_{\alpha \beta} F^{\beta \nu} + \frac{1}{4} \eta^{\mu \nu} F^{\alpha \beta} F_{\alpha \beta} ) $ is the electromagnetic stress-energy tensor. If written out in full we find
\be \label{EMStressEnergyMatrix}
T^{\mu \nu } = \begin{bmatrix} u & S_x/c & S_y/c & S_z/c \\
S_x/c & u - T_{xx}  & T_{xy} & T_{xz} \\
S_y/c & T_{yx} & u - T_{yy} & T_{yz} \\
S_z/c & T_{zx} & T_{zy} & u - T_{zz}\end{bmatrix} 
\ee
where $ T_{ij} = \epsilon_0 E_i E_j + \frac{1}{\mu_0} B_i B_j $.
We note from Eq.~(\ref{FeoFExpansion}), that the first row $ T^{0 \nu } $ is given by $ \frac{1}{2} \epsilon_0 F e_0 F $.  In fact the next three rows can be shown to be 
\be
T_i = F e_i e_0 F ,
\ee
where $ i \in 1 \dots 3$.  Hence the complete stress energy tensor is given by  \cite{Hestenes:1966}
\be \label{eq:StressEnergyTensorGAe0}
T^{\mu \nu} \leftrightarrow F \underline{n} F ,
\ee
where $  \underline{n} = (1 + e_1 +e_2 +e_3) e_0  \leftrightarrow n^{\mu } $ or explicitly $ T^{\mu \nu} = (F n^{\mu} F) \cdot n^{\nu} $.
Therefore we can now write Eq.~(\ref{MomentumConservationTensor}) using GA as
\be \label{GAMomentumConservation}
\Box \cdot T_i  = k_i 
\ee
for the three equations of momentum conservation, where the vector force components are given by Eq.~(\ref{MinkowskiForce}).

\begin{acknowledgments}
Thanks are due to Withawat Withayachumnankul for assistance with the diagrams.
\end{acknowledgments}

\end{document}